\documentclass[twocolumn,superscriptaddress,showpacs,prd,aps,amsmath,amssymb,nofootinbib]{revtex4-1}
\usepackage{graphicx}
\usepackage{amssymb}
\usepackage{bm}
\usepackage{color}

\newcommand{\Madm}{M_{\rm ADM}}
\newcommand{\MK}{M_{\rm K}}
\newcommand{\Mirr}{M_{\rm irr}}
\newcommand{\AHarea}{A_{\rm AH}}
\newcommand{\beq}{\begin{equation}} 
\newcommand{\eeq}{\end{equation}} 
\newcommand{\beqn}{\begin{eqnarray}} 
\newcommand{\eeqn}{\end{eqnarray}} 
\newcommand{\pa}{\partial}
\newcommand{\na}{\nabla}

\newcommand{\gabu}{g^{\alpha\beta}}

\newcommand{\gmabu}{\gamma^{ab}}

\newcommand{\albe}{{\alpha\beta}}

\newcommand{\Gabd}{G_{\alpha\beta}}

\newcommand{\Kab}{K^a{}\!_b}

\newcommand{\zD}{{\raise1.0ex\hbox{${}^{\ \circ}$}}\!\!\!\!\!D}
\newcommand{\alone}{{\raise0.5ex\hbox{${}^{\ 1}$}}\!\!\!\!\alpha}
\newcommand{\Od}{{O}}

\newcommand{\dl}{\delta}

\newcommand{\Dl}{\Delta}

\newcommand{\Lie}{\mbox{\pounds}}

\newcommand{\nalam}{\mathrel{\raise0.9ex\hbox{$^\lambda$}\mkern-14mu
\lower0.0ex\hbox{$\nabla$}}}

\newcommand{\Nrf}{{N_r^{\rm f}}}
\newcommand{\Nrm}{{N_r^{\rm m}}}

\newcommand{\Kabd}{K_{ab}}

\newcommand{\zeroD}{{\raise1.0ex\hbox{${}^{\ \circ}$}}\!\!\!\!\!D}
\newcommand{\Lap}{\Delta}
\newcommand{\zLap}{{\raise1.0ex\hbox{${}^{\ \circ}$}}\!\!\!\!\Delta}
\newcommand{\zna}{{\raise1.0ex\hbox{${}^{\ \circ}$}}\!\!\!\!\!\nabla}
\newcommand{\zS}{{\raise1.0ex\hbox{${}^{\ \circ}$}}\!\!\!\!\!S}

\newcommand{\cocal}{{\sc cocal }}

\newcommand{\kadath}{{\tt KADATH }}

\newcommand{\GA}{\alpha}
\newcommand{\GB}{\beta}
\newcommand{\GG}{\gamma}

\newcommand{\GC}{\psi}

\newcommand{\pd}{\partial}

\newcommand{\be}{\begin{equation}}
\newcommand{\ee}{\end{equation}}

\begin{document}

\title{
New code for equilibriums and quasiequilibrium initial data of compact objects. II. 
Convergence tests and comparisons of binary black hole initial data}

\author{K\=oji Ury\=u}
\email{uryu@sci.u-ryukyu.ac.jp}
\affiliation{Department of Physics, University of the Ryukyus, Senbaru, Nishihara, 
Okinawa 903-0213, Japan}

\author{Antonios Tsokaros}
\email{atsok@aegean.gr}
\affiliation{Department of I.C.S.E., University of Aegean, Karlovassi 83200, Samos, Greece}

\author{Philippe Grandcl\'ement}
\email{philippe.grandclement@obspm.fr}
\affiliation{Laboratoire Univers et Th\'eories, UMR 8102 du CNRS,
Observatoire de Paris, Universit\'e Paris Diderot, F-92190 Meudon, France}

\date{\today}

\begin{abstract}
{\sc Cocal} is a code for computing equilibriums 
or quasiequilibrium initial data of single or 
binary compact objects based on finite difference methods.  
We present the results of supplementary convergence tests 
of \cocal code using time symmetric binary black 
hole data (Brill-Lindquist solution).  
Then, we compare the initial data 
of binary black holes on the conformally flat spatial 
slice obtained from \cocal and {\tt KADATH}, 
where \kadath is a library for solving a wide class 
of problems in theoretical physics including 
relativistic compact objects with spectral methods.  
Data calculated from the two codes converge 
nicely towards each other, for close as well as largely separated 
circular orbits of binary black holes. 
Finally, as an example, a sequence of equal mass 
binary black hole initial data with corotating spins 
is calculated and compared with data in the literature.  
\end{abstract}

\maketitle

\section{Introduction}
\label{sec:int}

Various methods 
have been developed in the past couple of decades 
for computing numerical solutions of 
compact objects in equilibrium or quasiequilibrium.  
Those include methods for 
computing relativistic rotating stars in equilibrium 
or binary black hole initial data (see e.g. 
\cite{RNSreview,Cook:2000vr}).  
For example, 
numerical solutions of binary black holes (BBH) 
in quasicircular orbits have been widely used for 
initial data of merger simulations \cite{Pfeiffer:2012pc,BBH_QE}, 
and sequences of such data with fixed irreducible 
mass of each black hole (BH) have been also applied to approximate 
the inspiral evolution of BBH \cite{Caudill:2006hw}.

The authors have been developing, independently, 
numerical codes for computing such compact objects, 
\cocal \cite{Uryu:2011ky} (Paper I hereafter) and 
\kadath \cite{Grandclement:2009ju} (Paper II). 
{\sc Cocal} is a code for computing various kinds 
of astrophysical compact objects - isolated or binary 
systems of neutron stars and black holes which may 
be associated with strong magnetic fields.  
\kadath is a library for solving a wide class 
of problems in theoretical physics including those of 
general relativity, and is capable of computing 
such compact objects.

In the first part of this paper, we present 
the results of supplementary convergence tests of \cocal 
to those presented in Paper I.  With straightforward 
changes in the radial coordinate grid spacings and 
in the finite difference formula for the integration 
over the zenith angle, errors in the gravitational 
fields especially near the compact objects decrease 
substantially, which is necessary to improve the 
accuracy of widely separated BBH solutions.  
In the second part of the paper, 
we carefully compare the spatially conformally flat 
BBH initial data in circular orbit calculated from 
\cocal and \kadath code.  
Comparison of the solutions is the most effective 
test to confirm the reliability of the codes in which a 
system of complicated equations is solved.  
Such comparison had been done for the codes 
for relativistic rotating stars in \cite{Nozawa:1998ak}.  
To our knowledge, this is the 
first attempt to compare the BBH data 
calculated from totally different methods as 
the spectral method and the finite difference method. 
Finally, we present a sequence of 
spatially conformally flat BBH initial data in circular 
orbits for the case with equal mass and corotating spin, 
and compare the result with those presented in 
\cite{Caudill:2006hw}.  Throughout the paper 
we use geometric units with $G=c=1$.

\section{Convergence tests for \cocal code}
\label{sec:convtest}

In this section, we present convergence tests of 
\cocal code supplementary to those presented in 
Paper I.  The setup for the test problem is the 
same as in Paper I: the Brill-Lindquist 
solution for the time symmetric BBH data 
is generated numerically, and it is compared 
with the analytic value.  We briefly review the setup of 
the problem and discuss the modification of the finite 
difference scheme from the previous paper.  

\subsection{A test problem for binary black holes}

We assume the spacetime $\cal M$ is foliated 
by a family of spacelike hypersurfaces 
$(\Sigma_t)_{t\in {\mathbb R}}$, 
${\cal M} = {\mathbb R} \times \Sigma$ parametrized by 
$t\in {\mathbb R}$.  
We assume the line element in the neighborhood of $\Sigma_t$
to be 
\be
ds^{2}=-\GA^{2}dt^{2}+\GC^{4}f_{ij}dx^{i}dx^{j} \:,
\ee
where $f_{ij}$ is the flat spatial metric, 
so that the data on $\Sigma_t$ becomes time symmetric - 
the extrinsic curvature $\Kabd$ on $\Sigma_t$ vanishes.

Decomposing Einstein's 
equation $\Gabd=0$ with respect to the foliation using 
hypersurface normal $n^\alpha$ to $\Sigma_t$, and the projection 
tensor $\gmabu = \gabu + n^\alpha n^\beta$ to it, 
we write the Hamiltonian constraint 
$\Gabd n^\alpha n^\beta=0$, and a combination of the spatial trace of 
Einstein's equation and the constraint 
$G_{\GA\GB}(\gamma^{\GA\GB} + \frac12n^\alpha n^\beta)=0$, as 
\be
\nabla^{2}\GC=0 \:\:\:\:\:\: \textrm{and} \:\:\:\:\:\: \nabla^{2}(\GA\GC)=0 \: .
\label{eq:Laplace}
\ee 
These equations have solutions, which correspond to the Schwarzschild metric 
in isotropic coordinates for a single BH.  For a two BH case, a BBH solution 
is given by Brill and Lindquist \cite{BL63}:  
\beq
\GC \,=\, 1+\frac{M_1}{2r_1}+\frac{M_2}{2r_2}
\:\:\:\: \textrm{and} \:\:\:\:
\alpha\psi \,=\, 1-\frac{M_1}{2r_1}-\frac{M_2}{2r_2}, 
\label{eq:psi_alps_2bhsol}
\eeq
where subscripts 1 and 2 corresponds to those of the first and second BH; 
$r_1$ and $r_2$ are distances from the first and second BH, respectively, 
and $M_1$ and $M_2$ are mass parameters.  
Instead of solving two Laplace equations Eq.~(\ref{eq:Laplace}), we write 
an equation for $\alpha$ with a source on the whole domain of $\Sigma_t$: 
\be
\nabla^{2}\GC=0 \:\:\:\:\:\: \textrm{and} \:\:\:\:\:\: 
\nabla^{2}\GA=-\frac{2}{\GC}f^{ij}\pd_i\GC \pd_j\GA .
\label{eq:Laplace2}
\ee 
In an actual computation, spherical regions near the center of BH 
are excised to avoid singularities.  Therefore, 
boundary conditions for these elliptic equations are imposed 
at the radius $r=r_a$ 
of the excised sphere $S_a$, and at the radius $r=r_b$ of 
the boundary of computational domain $S_b$.  
We also set the mass parameters $M_1$ and $M_2$ as 
$0.8 \times r_a$ of each BH to avoid the lapse to be negative 
at $S_a$.  
In the following tests, we impose Dirichlet boundary conditions 
on $S_a$ and $S_b$ whose values are 
taken from the analytic solution (\ref{eq:psi_alps_2bhsol}).

\subsection{Coordinates, grid setup, and finite difference scheme 
of \cocal code}
\label{sec:GridSp}

As explained in Paper I, three spherical coordinate 
patches are introduced for solving binary compact objects 
with \cocal.  Two of them are the compact objects coordinate 
patches (COCP-I and II) and one is the asymptotic region 
coordinate patch (ARCP).  
In each spherical patch, coordinates cover the region $(r,\theta,\phi) 
\in [r_a,r_b]\times[0,\pi]\times[0,2\pi]$.  
The two COCPs are centered at the center of 
compact objects and extend up to about $r_b\sim\Od(10^2 M)$, while 
ARCP is centered at the center of mass of the binary, and 
extends from $r_a\sim\Od(10M)$ to $r_b\sim\Od(10^6 M)$, 
where $M$ is the total mass of the system.  
Definitions of the parameters for the grid setups are 
listed in Table \ref{tab:grids_param}.

In solving a system of elliptic equations such as Eq.(\ref{eq:Laplace2}), 
we rewrite them in integral form using Green's 
function that satisfies given boundary conditions, 
and apply a finite difference scheme to discretize those integral 
equations on the spherical coordinates of each domain.  
We use the midpoint rule for numerical quadrature formula, 
and hence compute the source terms at the midpoints of the grids.  
Coordinate grids $(r_i,\theta_j,\phi_k)$ with 
$i = 0, \cdots, N_r$, $j = 0, \cdots, N_\theta$, 
and $k = 0, \cdots, N_\phi$, are freely specifiable except for the
endpoint of each coordinate grid that corresponds to the boundary of 
the computational region, 
$(r_0,\theta_0,\phi_0) = (r_a,0,0)$ 
and 
$(r_{N_r},\theta_{N_\theta},\phi_{N_\phi}) = (r_b,\pi,2\pi)$. 
The grid setup for COCP and ARCP is the same as Paper I 
except for the radial grid of COCP which will be explained later.  
For angular coordinate grids $(\theta_j,\phi_k)$, 
we choose equally spaced grids.

In Paper I, 
we have used for the finite difference formulas, 
(1) 2nd order midpoint rule for the quadrature formula, 
(2) 2nd order finite difference formula for the $\theta$ 
and $\phi$ derivatives evaluated at the mid points 
$(r_{i+\frac12},\theta_{j+\frac12},\phi_{k+\frac12})
=((r_{i}+r_{i+1})/2,(\theta_{i}+\theta_{i+1})/2,
(\phi_{i}+\phi_{i+1})/2)$, 
(3) 3rd order finite difference formula for the $r$ 
derivative evaluated at the mid points, and 
(4) 4th order finite difference formula for the 
derivatives evaluated at the grid points, 
$(r_{i},\theta_{j},\phi_{k})$.  
In the present computations, we use the same finite difference 
formulas mentioned above except for the numerical quadrature 
formula in $\theta$ integrations.  

Differences from the previous Paper I are 
the spacings of radial grids $\Dl r_i := r_i - r_{i-1}$ in COCP, 
and the quadrature formula used for the integration 
in zenith angle $\theta$.  

\begin{table}
\begin{tabular}{lll}
\hline
$r_a$  &:& Radial coordinate where the radial grids start.       \\
$r_b$ &:& Radial coordinate where the radial grids end.     \\
$r_c$ &:& Radial coordinate between $r_a$ and $r_b$ where   \\
&\phantom{:}& the radial grid spacing changes.   \\
$r_e$  &:& Radius of the excised sphere.       \\
$N_{r}$ &:& Number of intervals $\Dl r_i$ in $r \in[r_a,r_{b}]$. \\
$\Nrf$ &:& Number of intervals $\Dl r_i$ in $r \in[r_a,1]$. \\
$\Nrm$ &:& Number of intervals $\Dl r_i$ in $r \in[r_a,r_{c}]$. \\
$N_{\theta}$ &:& Number of intervals $\Dl \theta_j$ in $\theta\in[0,\pi]$. \\
$N_{\phi}$ &:& Number of intervals $\Dl \phi_k$ in $\phi\in[0,2\pi]$. \\
$d$ &:& Coordinate distance between the center of $S_a$ ($r=0$) \\
&\phantom{:}& and the center of mass. \\
$d_s$ &:& Coordinate distance between the center of $S_a$ ($r=0$) \\
&\phantom{:}& and the center of $S_e$. \\
$L$ &:& Order of included multipoles. \\
\hline
\end{tabular}  
\caption{Summary of grid parameters for COCP. 
}
\label{tab:grids_param}
\end{table}
\begin{figure}
\begin{center}
\includegraphics[height=60mm]{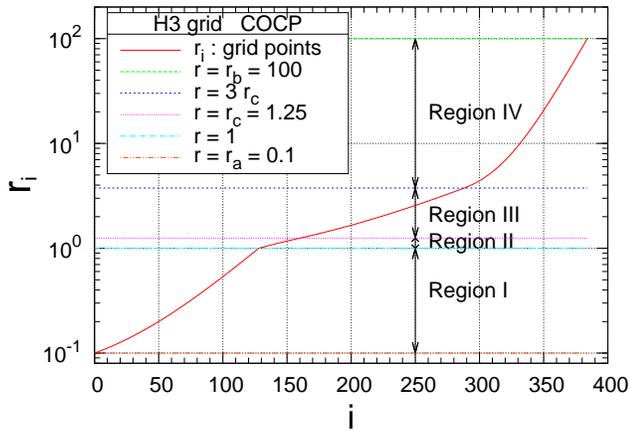}
\caption{The radial grid points $r_i$ of COCP are plotted 
against the grid number $i = 0, \cdots, N_r$ 
for the case with H3 grid in Table \ref{tab:BBHtest_grids}.  
}
\label{fig:radial_grid}
\end{center}
\end{figure}

\subsubsection{Radial grid spacings for COCP}
\label{sec:grid_r_bhex}

When we compute a sequence of BBH data from 
larger to smaller separations in the \cocal code, 
we change the BH excision radius $r_a$ from smaller 
to larger values and fix the separation $d_s$ 
(instead of fixing $r_a$ and varying $d_s$).  
In this way, the number of grid points are kept 
to be the same, and the structures of coordinate grids 
are almost the same for all solutions of the sequence.  
As a result the discretization error behaves 
systematically from one solution to the other, and hence 
the quantities such as mass or angular momentum 
vary smoothly along a sequence of solutions.

It is important to notice that, by changing the 
BH radius, we change the mass of the solution 
and hence change the length scale of the system.  
Therefore, to maintain the accuracy of the gravitational 
fields near the BH, the intervals 
near the hole should be proportional to the mass of 
the BH, or in our case, the BH excision radius $r_a$.  
Therefore, we 
modify the construction of the grid spacing in 
the radial direction $r$ of COCP for computing 
a sequence from smaller to larger BH as follows.  
Without loss of generality, we set the radius of BH excision 
sphere $S_a$ as $r_a < 1$.  We divide the radial coordinate 
to four regions, I: $r\in[r_a,1]$, II: $r\in[1,r_c]$, 
III: $r\in[r_c,3r_c]$, and IV: $r\in[3r_c,r_b]$.   
We set the first interval by 
\beq
\Delta r_1 = \frac{r_a}{\lambda\Nrf}, 
\eeq
where $\Nrf$ is the number of intervals in the 
region I:$[r_a,1]$, and $\lambda$ is a constant factor 
which is chosen to be $\lambda=0.75$.  
For each region, $\Dl r_i := r_i - r_{i-1}$, are defined by 
\beqn
\Dl r_{i+1} &=& h_1 \Dl r_i,\ \ \mbox{for}\ \  i = 1, \cdots, \Nrf-1
\\
\Dl r_{i\phantom{+1}} &=& \phantom{h_1} \Dl r,\ \ \ \mbox{for}\ \  i = \Nrf, \cdots, \Nrm
\\
\Dl r_{i+1} &=& h_3 \Dl r_i,\ \ \mbox{for}\ \  i = \Nrm, \cdots, \Nrm+\Nrf-1 \ \ 
\\
\Dl r_{i+1} &=& h_4 \Dl r_i,\ \ \mbox{for}\ \  i = \Nrm+\Nrf, \cdots, N_r-1 \ \ 
\eeqn
which correspond to regions I, II, III, and IV, respectively, 
The ratios $h_i(> 1)$ $(i=1,3,4)$ are respectively determined from relations
\beqn
1-r_a &=& \Dl r_1 \frac{h_1^\Nrf-1}{h_1-1}, 
\\
2 r_c &=& \Dl r \frac{h_3(h_3^{\Nrf}-1)}{h_3-1}.  
\\
r_b - 3r_c &=& \Dl r \frac{h_4(h_4^{N_r-\Nrm-\Nrf}-1)}{h_4-1}.  
\label{eq:coord_r_ratio_k}
\eeqn

Values of the parameters for the coordinate grids of \cocal 
used in computing the results presented in this paper 
are listed in Table \ref{tab:BBHtest_grids}.
In Fig.~\ref{fig:radial_grid}, an example of the radial grid points 
is plotted for the case with H3 grid setup in Table \ref{tab:BBHtest_grids}.  
Because of the construction, the grid structure 
in the region larger than $r \ge 1$ is the same 
for all solutions with different BH radius $r_a$ 
once a grid setup (resolution) as 
in Table \ref{tab:BBHtest_grids} is selected.

\subsubsection{4th order midpoint rule for 
the quadrature formula of $\theta$ integration}
\label{sec:th4th}

As discussed in Paper I, our Poisson solver 
is a system of integral equations, and 
it is numerically integrated with 
a quadrature formula of midpoint rule.  
Therefore, the sources of the integrals are always 
evaluated at the midpoints of ($r_i,\theta_j,\phi_k$) 
grids.  As summarized above, the 2nd order midpoint 
rule was used for a quadrature formula in Paper I.  
With the above mentioned choice for finite difference formulas, 
the 2nd order convergence of the error has been achieved.  
As shown in the top panel of Fig.~\ref{fig:noneqmBBHtest_ra02} 
(as well as figures 3, 4, and 6 in Paper I), 
however, the fractional error of the potentials 
normally increases near the excision surfaces of the BH, 
$S_a$ $(r=r_a)$, although it converges in 2nd order.  

One might expect that the increase of the error near $S_a$ 
is due only to a lack of resolution in radial grid points.  
It turns out, however, that the finite difference errors 
in the potentials near the boundaries of computational domains 
are dominated by the discretization error in the $\theta$ coordinate.  
In particular, 
the $\theta$ integration of the source involving 
the Legendre function turns out to be the source of error.  
Therefore, we replace the quadrature formula of $\theta$ 
integration to 4th order accurate midpoint rule
whose weights are 
\beqn
&& 
\int_{\theta_j}^{\theta_{j+4}} f(\theta) d\theta 
\,\simeq\,
\nonumber\\
&& \quad 
\Delta\theta\left(\frac{13}{12}f_{j+\frac12} + 
\frac{11}{12}f_{j+\frac32}
+ \frac{11}{12}f_{j+\frac52} 
+ \frac{13}{12}f_{j+\frac72}\right), 
\label{eq:th4th}
\eeqn
where the grid number $j$ is a multiple of 4, and 
$\Delta \theta = \pi/N_\theta$.  
%

%
%
\begin{table}
\begin{tabular}{clccccrrrrrrr}
\hline
Type & Patch & $r_a$ & $r_b$ & $r_c$ & $r_e$ & $\Nrf$ & $\Nrm$ & $N_r$ & $N_\theta$ & $N_\phi$ & $L$ \\
\hline
H1 & COCP-1 & var. & $10^2$ & 1.25 & 1.125 & 32  & 40   & 96   & 24  & 24 &  12 \\
   & COCP-2 & var. & $10^2$ & 1.25 & 1.125 & 32  & 40   & 96   & 24  & 24 &  12 \\
   & ARCP   & 5.0 & $10^6$ & 6.25 &  ---  &  8  & 10   & 96   & 24  & 24 &  12 \\
H2 & COCP-1 & var. & $10^2$ & 1.25 & 1.125 & 64  & 80   & 192  & 48  & 48 &  12 \\
   & COCP-2 & var. & $10^2$ & 1.25 & 1.125 & 64  & 80   & 192  & 48  & 48 &  12 \\
   & ARCP   & 5.0 & $10^6$ & 6.25 &  ---  & 16  & 20   & 192  & 48  & 48 &  12 \\
H3 & COCP-1 & var. & $10^2$ & 1.25 & 1.125 & 128 & 160  & 384  & 96  & 96 &  12 \\
   & COCP-2 & var. & $10^2$ & 1.25 & 1.125 & 128 & 160  & 384  & 96  & 96 &  12 \\
   & ARCP   & 5.0 & $10^6$ & 6.25 &  ---  & 32  & 40   & 384  & 96  & 96 &  12 \\
H4 & COCP-1 & var. & $10^2$ & 1.25 & 1.125 & 256 & 320  & 768  & 192  & 192 &  12 \\
   & COCP-2 & var. & $10^2$ & 1.25 & 1.125 & 256 & 320  & 768  & 192  & 192 &  12 \\
   & ARCP   & 5.0 & $10^6$ & 6.25 &  ---  & 64  & 80   & 768  & 192  & 192 &  12 \\
\hline
\end{tabular}
\caption{Grid parameters of \cocal 
used for computation of BBH data.  
The separation of two BHs is fixed as $d_s=2.5$.
For the excision radius $r_a$ for COCP-1 and 2, 
``var.'' stands for a variable parameter 
assigned to each solution.  In the test problems 
in Sec.\ref{sec:conv_test}, they are chosen to be 
$r_a=0.2$ and $0.4$ for close BBH, 
and $r_a=0.05$ and $0.1$ for separated BBH
for COCP-1 and 2, respectively.}
\label{tab:BBHtest_grids}
\end{table}

\subsection{Convergence tests}
\label{sec:conv_test}

We perform convergence tests to examine that the 
above two modifications improve the accuracy of 
the \cocal code.  
In Table \ref{tab:BBHtest_grids}, grid setups for the 
computations are listed.  The grids H1--H4 
correspond to different levels of resolutions.  
At each level, the resolution is double 
the previous one\footnote{
We have also tested different combinations of grid numbers 
$(N_r,N_\theta,N_\phi)$ for the first level of resolution 
and performed convergence tests.  We found 
the combination of type H was better than others.  
For example, the accuracy was not improved by 
increasing the grid points in the $\phi$ direction.}.

\begin{figure}
\begin{center}
\includegraphics[height=60mm]{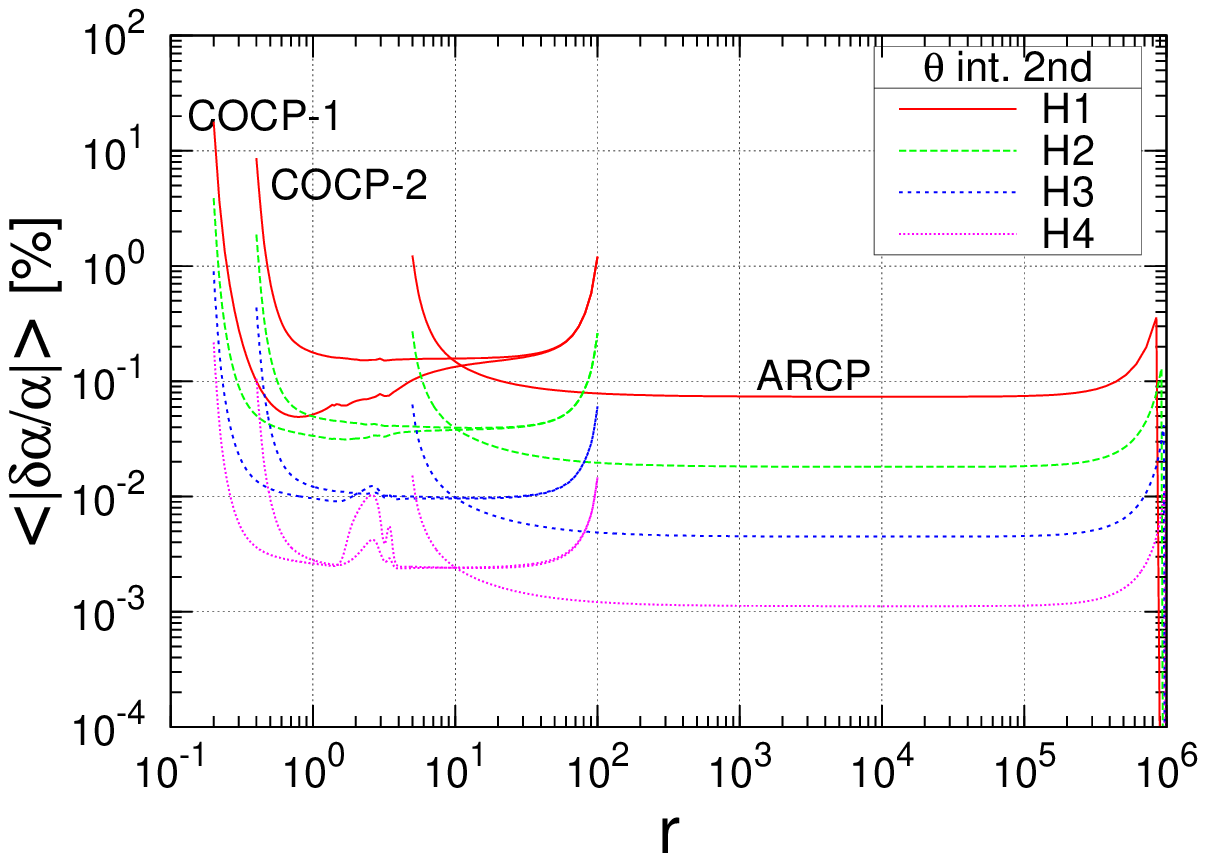}
\includegraphics[height=60mm]{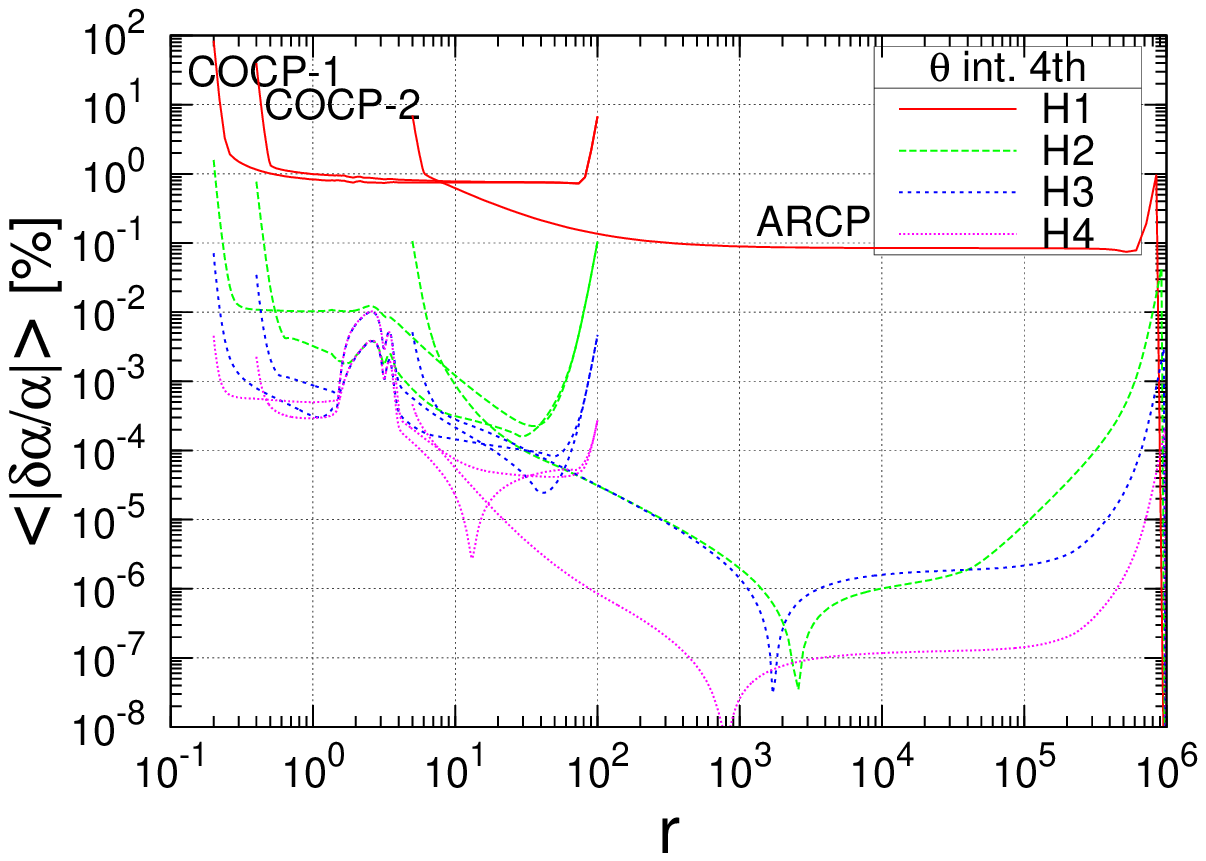}
\caption{Angular averaged fractional errors in the lapse 
$\left< \left| \dl \alpha/\alpha \right| \right>$ are plotted 
along the radial coordinate $r$ for the nonequal mass 
BBH data computed on three multiple patches.  
Top panel: data computed 
with the same coordinate grid spacings and finite difference 
schemes as in Paper I.  
Bottom panel: data computed with the 4th order 
$\theta$ integration discussed in Sec.\ref{sec:th4th}.
The grid parameters and number of grid points are varied 
as H1-H4 in Table \ref{tab:BBHtest_grids}.
The BH excision radii are chosen to be $r_a = 0.2$ and $0.4$ for 
COCP-1 and 2, respectively, and the separation is to be $d_s=2.5$.  
}
\label{fig:noneqmBBHtest_ra02}
\end{center}
\end{figure}
\begin{figure}
\begin{center}
\includegraphics[height=60mm]{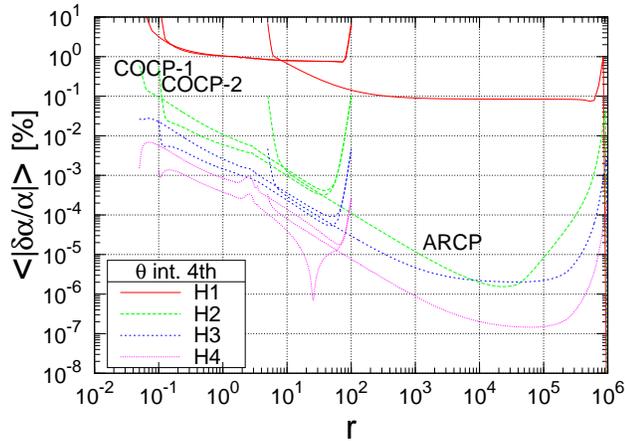}
\caption{Same as Fig.~\ref{fig:noneqmBBHtest_ra02}, 
but for the BH radius $r_a = 0.05$ and $0.1$ 
for COCP-1 and 2, respectively.  
The results are calculated using the 4th order midpoint 
rule for $\theta$ integration as in the bottom panel of 
Fig.\ref{fig:noneqmBBHtest_ra02}, otherwise the same 
finite differencing scheme (in particular, the same 
radial grid spacing $\Dl r_i$) as in Paper I.  
}
\label{fig:noneqmBBHtest_ra005}
\end{center}
\end{figure}

\begin{figure}
\begin{center}
\includegraphics[height=60mm]{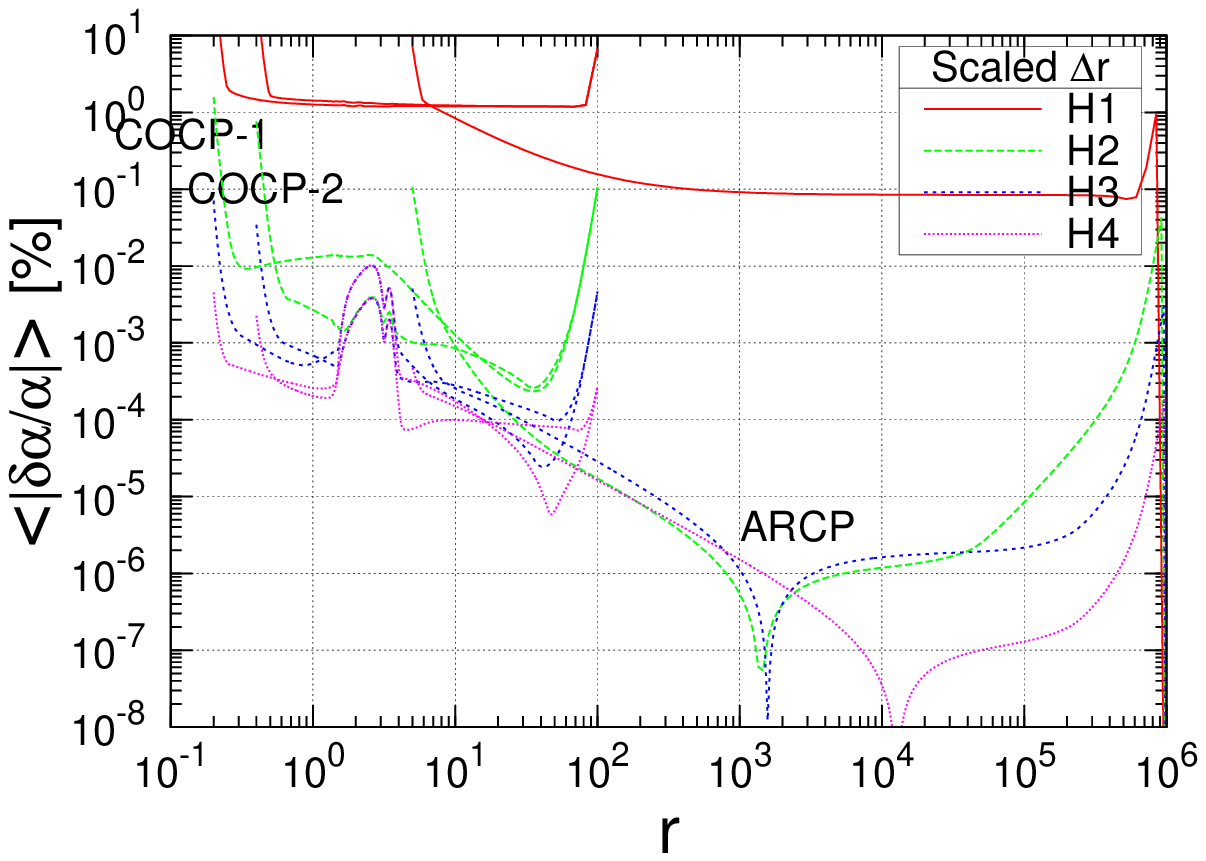}
\includegraphics[height=60mm]{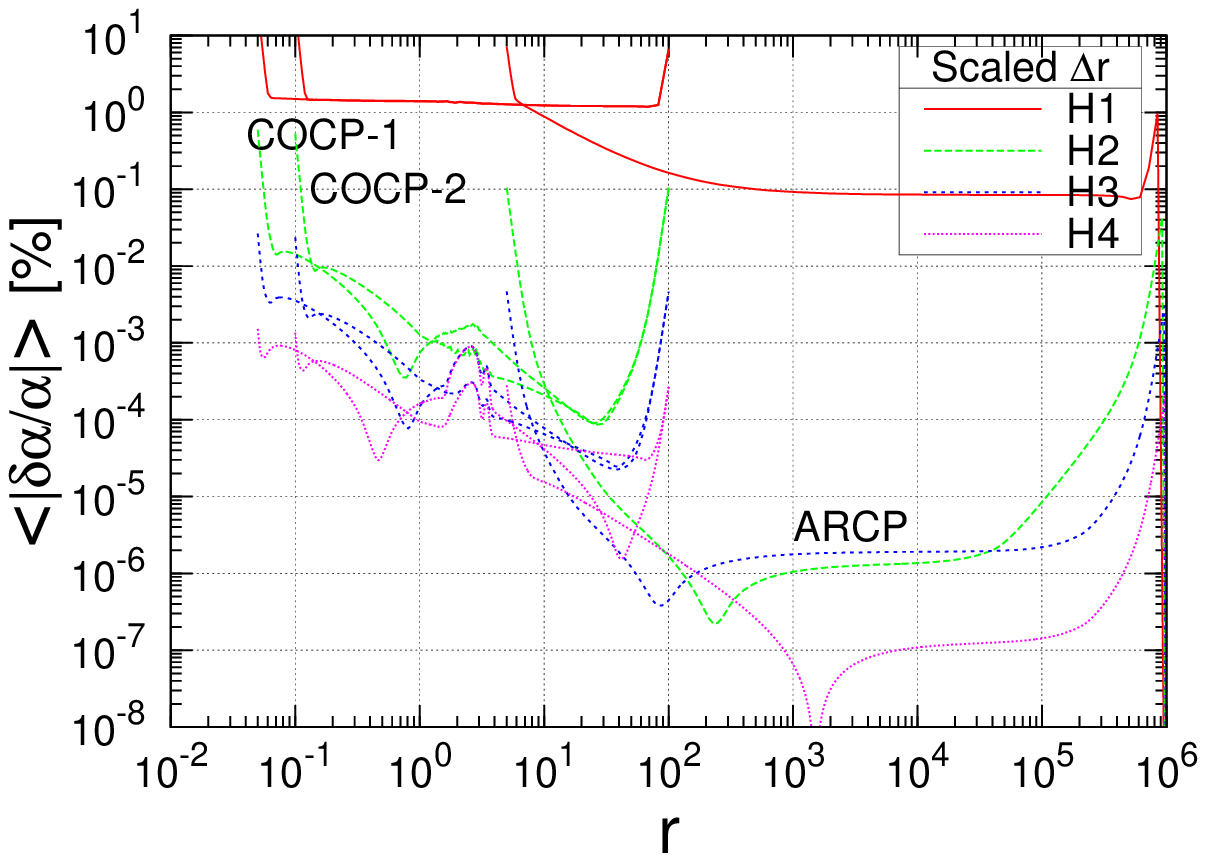}
\caption{Same as Fig.~\ref{fig:noneqmBBHtest_ra02}.  
The results are calculated using 4th order integration in 
$\theta$ coordinate Eq.~(\ref{eq:th4th}), and the scaled 
radial spacing discussed in Sec.\ref{sec:grid_r_bhex}.  
Top panel: data with the BH radius $r_a = 0.2$ and $0.4$ for 
COCP-1 and 2, respectively.  
Bottom panel: data with the BH radius $r_a = 0.05$ and $0.1$ for 
COCP-1 and 2, respectively.  
}
\label{fig:noneqmBBHtest_grid_r_bhex}
\end{center}
\end{figure}

In Figs.~\ref{fig:noneqmBBHtest_ra02}--\ref{fig:noneqmBBHtest_grid_r_bhex}, 
the fractional errors in the lapse that are averaged over the 
angular coordinate grids $(\theta_i,\phi_j)$ 
at fixed radial coordinate $r$ are plotted against $r$ 
of each coordinate patch, 
\beq
\left< \left| \frac{\dl \alpha}{\alpha} \right| \right>
:= \frac1{\#({\cal G}_i)} \sum_{p \in {\cal G}_i}
\left|\frac{\alpha - \alpha_{\rm exact}}{ \alpha_{\rm exact}}\right| , 
\label{eq:frac_error_ave}
\eeq
where writing a grid point $(r_i,\theta_j,\phi_k)$ by $p$, 
we define a set ${\cal G}_i$ by 
${\cal G}_i := \left\{p \ | \  
p\in V\setminus S^{\rm in}_e \ \mbox{and}\  r_i = \mbox{const} \right\}$, 
where  $S^{\rm in}_e$ is an interior domain of $S_e$, 
and $\#({\cal G}_i)$ is the number of points included in ${\cal G}_i$.

In the top panel of Fig.\ref{fig:noneqmBBHtest_ra02}, 
the same finite difference scheme as presented in Paper I 
is used for computing closer BBH solution with 
the BH excision radius 
$r_a=0.2$ and $0.4$, and with the separation $d_s=2.5$.  
The midpoint rule in the $\theta$ integration is 2nd order 
accurate in this panel.  In the bottom panel, 
the 4th order midpoint rule Eq.(\ref{eq:th4th}) for 
the $\theta$ integration is used for the same model 
and the same grid spacings.  
Clearly, the 
fractional error substantially decreases by this change 
for the H2 to H4 levels.  We notice that the error 
near the BH converges in 4th order, that is, the error 
decreases about 1/16 at each level of resolution.  
The errors near the BH as well as in the asymptotic 
region are dominated by those from 
the $\theta$ integrations of the surface integral terms.

In Fig.\ref{fig:noneqmBBHtest_ra005}, we calculate more 
separated BBH solutions, decreasing the radius of BH 
excision surface to 1/4 (therefore effectively separating 
BBH 4 times apart) as $r_a = 0.05$ and $0.1$ with the 
same separation $d_s=2.5$, and using 
the 4th order integration in $\theta$ as in the bottom 
panel of Fig.\ref{fig:noneqmBBHtest_ra02}.  
Although the errors near the BH excision surfaces 
are of the same order of magnitude as those of the corresponding 
resolutions plotted in Fig.~\ref{fig:noneqmBBHtest_ra02}, 
bottom panel, the errors once 
increase as the radial coordinate $r$ increases.  
It turns out that the BBH initial data discussed later 
can not be calculated accurately with this grid setup 
for largely separated orbits.  

In Fig.\ref{fig:noneqmBBHtest_grid_r_bhex}, 
convergence tests for the close ($r_a = 0.2$ and $0.4$) 
and the separated ($r_a = 0.05$ and $0.1$) BBH are 
calculated with scaled radial spacing near the BH 
introduced in Sec.\ref{sec:grid_r_bhex}, 
as well as the 4th order midpoint rule in $\theta$ integration.  
The size of the errors around the BH excision radius 
for the largely separated BBH data (bottom panel) is now comparable 
to those for the close BBH data (top panel) for each level of resolution.  
This improvement turns out to be important for accurately computing the 
separated BBH data in \cocal code.

\section{Comparison of BBH initial data}

\subsection{The \kadath library}

In this section, we compare the circular solutions of 
BBH initial data on a conformally flat spacelike 
hypersurface calculated from \cocal and the \kadath library.  
\kadath \cite{Grandclement:2009ju} is a library designed to solve 
a wide class 
of problems in theoretical physics including those of 
general relativity such as the above compact objects.
It is based on spectral methods (see for instance \cite{lrr} and references therein)
where the various fields 
are approximated by finite sums of known functions typically
trigonometrical functions and orthogonal polynomials. One of the main
advantages of spectral methods is their fast convergence to the true 
solution (typically exponentially), when one increases the order of the expansion.
For instance, in this paper, a relative accuracy of about $10^{-4}$ is achieved with 15 
coefficients in each dimension. 

Spectral methods enable one to translate a set of partial differential equations into 
an algebraic system on the coefficients of the expansions. This system
is then solved by a standard Newton-Raphson iteration. The computation of 
the Jacobian as well as its inversion are parallelized.

The code used in this paper is essentially the same as the one used in
Sec.~7.3 of \cite{Grandclement:2009ju}. In order to check the overall 
accuracy of the computations, one monitors the convergence of some 
global quantities (like the orbital frequency), as a function
of $N$, the number of points in each dimensions. Let us mention that, 
in the case of a large separation, the code was slightly modified to maintain accuracy, 
probably due to a stretch of the bispherical coordinates when
 the distance between the holes gets much 
bigger than the size of the holes themselves. In particular, a spherical shell was added 
between the bispherical coordinates and the outer compactified domain and the 
determination of the orbital velocity had to be changed (see Sec. \ref{sec:comparison}).

\subsection{Conformally flat BBH initial data}

The circular solution of BBH initial data is calculated by 
solving the Hamiltonian and momentum constraints, and 
the spatial trace of the Einstein's equation on a 
conformally flat spacelike hypersurface $\Sigma_t$.  
The spacetime metric on $\Sigma_t$ is written in 3+1 form as
\beqn
ds^{2} & = & g_{\mu\nu}dx^{\mu}dx^{\nu} \nonumber \\
       & = & -\GA^{2}dt^{2}+\GG_{ij} (dx^{i}+\GB^{i}dt) (dx^{j}+\GB^{j}dt), 
\eeqn
where the spatial three metric $\GG_{ij}$ on the slice $\Sigma_t$ is assumed 
to be $\GG_{ij} = \GC^4 f_{ij}$.  
Here, field variables $\psi, \alpha$, and $\beta^i$ are the conformal factor, 
lapse, and shift vector, respectively, and $f_{ij}$ is a flat three 
dimensional metric.  We also assume maximal slicing 
to $\Sigma_t$, so that the trace of the extrinsic curvature 
$K_{ij} := -\frac1{2\alpha}( \Lie_t \GG_{ij}-\Lie_\beta \GG_{ij}) $ vanishes.  
Writing its tracefree part $A_{ij}$, 
the conformally rescaled quantity $\tilde A_{ij}$ becomes 
\beq
{\tilde A}_{ij} = 
\frac1{2\alpha} \left(\pd_i{\tilde \GB}_j + \pd_j{\tilde \GB}_i 
- \frac{2}{3}f_{ij}\pd_k{\tilde \GB}^k\right)\ ,
\eeq
where the derivative $\pa_i$ is associated with the flat metric 
$f_{ij}$, and conformally rescaled quantities with tilde are defined by 
$\tilde A_i{}^j= A_i{}^j$ and ${\tilde \beta}^i=\beta^i$, whose indexes 
are lowered (raised) by $f_{ij}$ ($f^{ij}$).  
The system to be solved, which are Hamiltonian and momentum 
constraints and the spatial trace of the Einstein's equation, becomes
\beqn
\Lap \GC &=& -\frac{\GC^5}{8}{\tilde A}_{ij}{\tilde A}^{ij},  
\label{eq:HC}\\
\Lap \GB_i &=& -2\,\alpha\, {\tilde A}_i{}^j \pd_j \ln\frac{\GC^6}{\GA}
-\frac{1}{3}\pd_i\pd_j{\tilde\GB}^j ,
\label{eq:MC}\\
\Lap (\GA\GC) &=& \frac{7}{8}\,\alpha\,\GC^5{\tilde A}_{ij}{\tilde A}^{ij},  
\label{eq:Kdot}
\eeqn
where $\Lap:=\pa_i\pa^i$ is a flat Laplacian \cite{Cook:2000vr,ISEN78,WM89}.

For the boundary conditions at the BH excision boundary 
$S_a$, we choose approximate irrotational apparent horizon boundary 
conditions, 
\beqn
\left.\frac{\pd\GC}{\pd r}+\frac{\GC}{2r}\right|_{r=r_a} 
& = & -\frac{\GC^3}{4}K_{ij}s^i s^j ,  
\label{eq:AHpsiBC} \\
\left.\GB^i\right|_{r=r_a} & = & \frac{n_0}{\GC^2}s^i 
+ \Omega\,y^i_{\rm cm},
\label{eq:AHbetaBC} \\ 
\left.\GA\right|_{r=r_a} & = & n_0  ,  
\label{eq:AHalphBC}
\eeqn
where $n_0$ is an arbitrary positive constant for which we choose $n_0 = 0.1$, 
$s^i$ is the unit normal to the sphere $S_a$, 
and $\Omega$ represents a parameter for orbital angular velocity.  
The vector $y^i_{\rm cm}:=(0,d,0)$ is the translational vector 
with respect to the center of mass.  
With these conditions, the sphere $S_a$ becomes an apparent horizon (AH) 
in quasiequilibrium \cite{Caudill:2006hw,Grandclement:2009ju,AHbcon}.

At the asymptotics, the boundary conditions are 
\beqn
& \left.\GC\right|_{r\rightarrow \infty} &=1.0  \\
& \left.\GB^i\right|_{r\rightarrow \infty} &=0.0  \\
& \left.\GA\right|_{r\rightarrow \infty} &=1.0.  
\label{eq:Initial_bc}
\eeqn
When using \kadath, the whole spacelike slice $\Sigma_t$ 
is compactified, and hence the above conditions are 
imposed at the spatial infinity, while in \cocal, 
the computational domain is truncated at the 
radius $r_b \sim \Od(10^6 M)$ and it is at 
$r_b$ that the above conditions are imposed.

\begin{table}
\begin{tabular}{lcllll}
\hline
Code & Res. & $\ \ \Omega \Mirr$ & $\Madm/\Mirr$ & $\ J/\Mirr^2$ & $\Mirr/r_a$   \\
\hline
 &  &  & $d_s/r_a=12$  &  &   \\
\kadath   & 11 & 0.127171 &\ 0.982866  & 0.757608 & 3.93366  \\
\kadath   & 13 & 0.127299 &\ 0.983041 & 0.758390 & 3.93405  \\
\kadath   & 15 & 0.127346 &\ 0.983072 & 0.758698 & 3.93417 \\
\cocal    & H2 & 0.127243 &\ 0.982848  & 0.754592 & 3.93191 \\
\cocal    & H3 & 0.127383 &\ 0.983125  & 0.758389 & 3.93392 \\
error [\%]& -- & 0.03 & 0.005 & 0.04 & 0.006 \\
\hline
 &  &  & $d_s/r_a=30$ &   &   \\
\kadath   & 11 & 0.0433696 &\ 0.933609 & 1.129241  & 3.56317  \\
\kadath   & 13 & 0.0350000 &\ 0.988279 & 0.906819 &  3.55178 \\
\kadath   & 15 & 0.0349704 &\ 0.988129 & 0.907249 &  3.55166  \\
\cocal    & H2 & 0.0344185 &\ 0.987369  & 0.889918 & 3.54574  \\
\cocal    & H3 & 0.0348624 &\ 0.987982  & 0.903226 & 3.55021  \\
error [\%]& -- & 0.3  & 0.01 & 0.4 & 0.04 \\
\hline
\end{tabular}
\caption{Comparison of BBH initial data from \cocal and \kadath for 
the cases $d_s/r_a = 12$ and $d_s/r_a = 30$.
Fractional differences in \% between the highest resolution cases of 
\kadath and \cocal are shown in the lines indicated by ``error''.  
A column "Res." stands for the resolution.}
\label{tab:BBH_comparison_D12D30}
\end{table}
\begin{figure}
\begin{tabular}{cc}
\begin{minipage}{.45\hsize}
\begin{center}
\includegraphics[height=60mm]{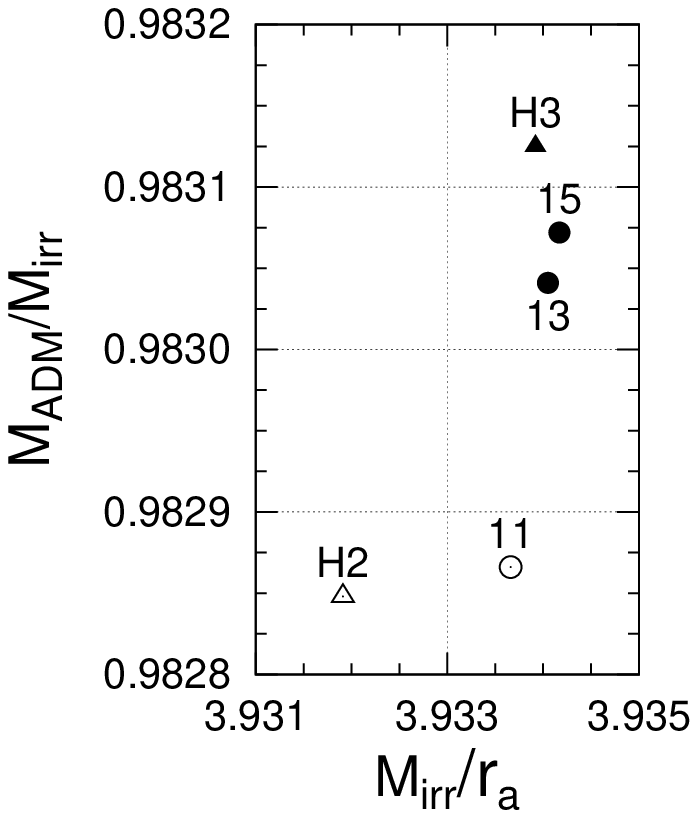}
\includegraphics[height=60mm]{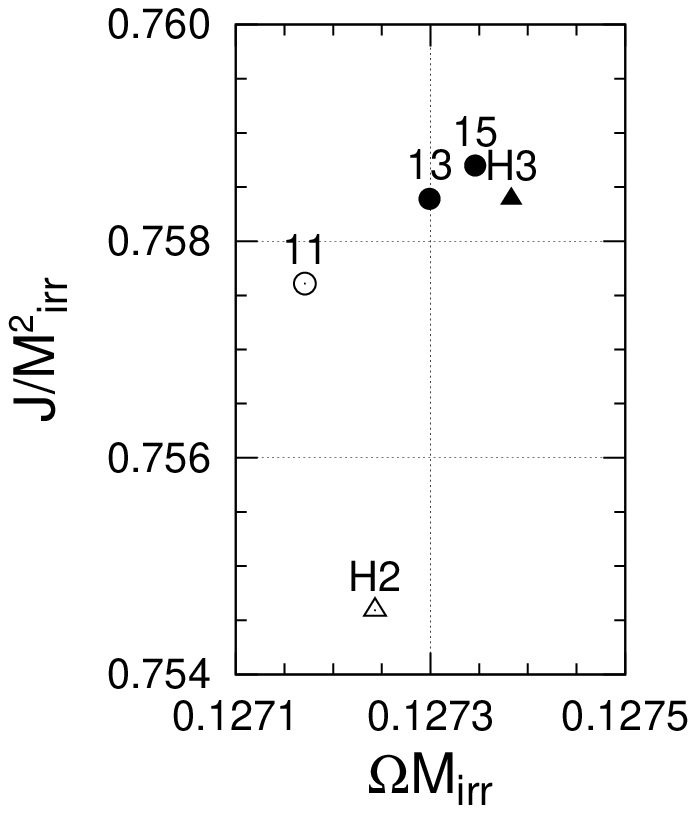}
\end{center}
\end{minipage}
&
\begin{minipage}{.55\hsize}
\begin{center}
\includegraphics[height=60mm]{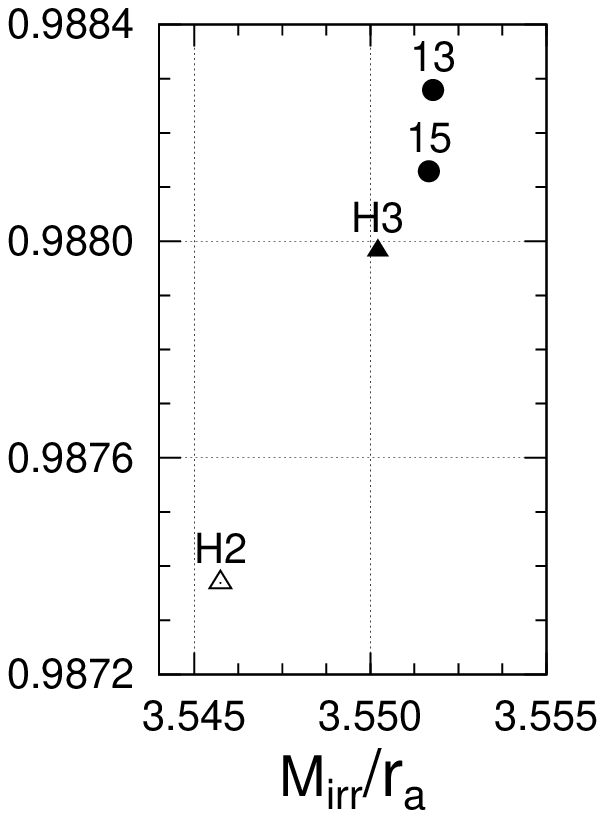}
\includegraphics[height=60mm]{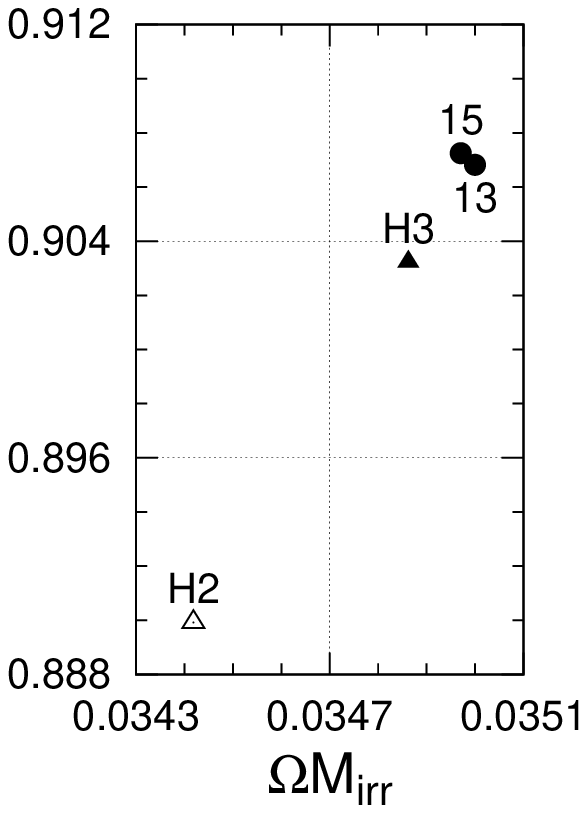}
\end{center}
\end{minipage}
\end{tabular}
\caption{
The panels show the scaled values of $\Madm/\Mirr$ as a function of $\Mirr/r_a$ and 
$J_{\rm ADM}/\Mirr^2$ as a function of $\Omega \Mirr$.
The circles denote the results from \kadath and the triangles the ones from \cocal. 
The runs using \kadath are labeled by the number of points in each dimension. 
The left top and bottom panels correspond to a separation of $d_s/r_a = 12$ and 
the right ones to $d_s/r_a = 30$ (in this case the lower 
resolution results from \kadath are outside the range of the plot).
}
\label{fig:conv}
\end{figure}
\begin{figure}
\begin{center}
\includegraphics[height=60mm]{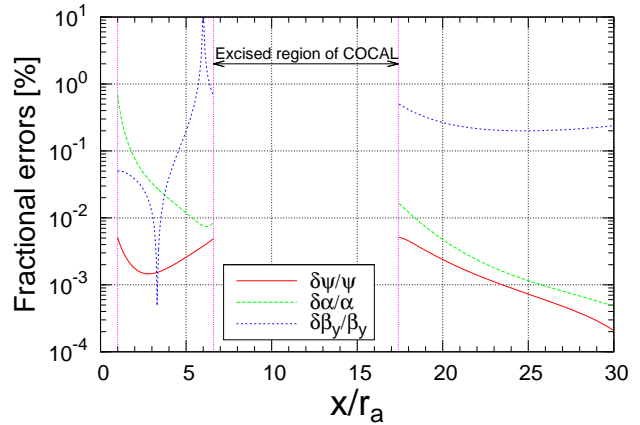}
\caption{Fractional errors in metric potentials 
between those calculated by \kadath and \cocal.  
Errors in the conformal factor $\psi$, lapse $\alpha$, 
and $y$-component of the shift $\beta_y$ are plotted 
along the $x$-axis that intersects with the centers of two BHs.  
The model is the same as that in Table \ref{tab:BBH_comparison_D12D30} 
with the separation $d_s/r_a=12$.  
}
\label{fig:plot_errors}
\end{center}
\end{figure}
\begin{figure}
\begin{center}
\includegraphics[height=60mm]{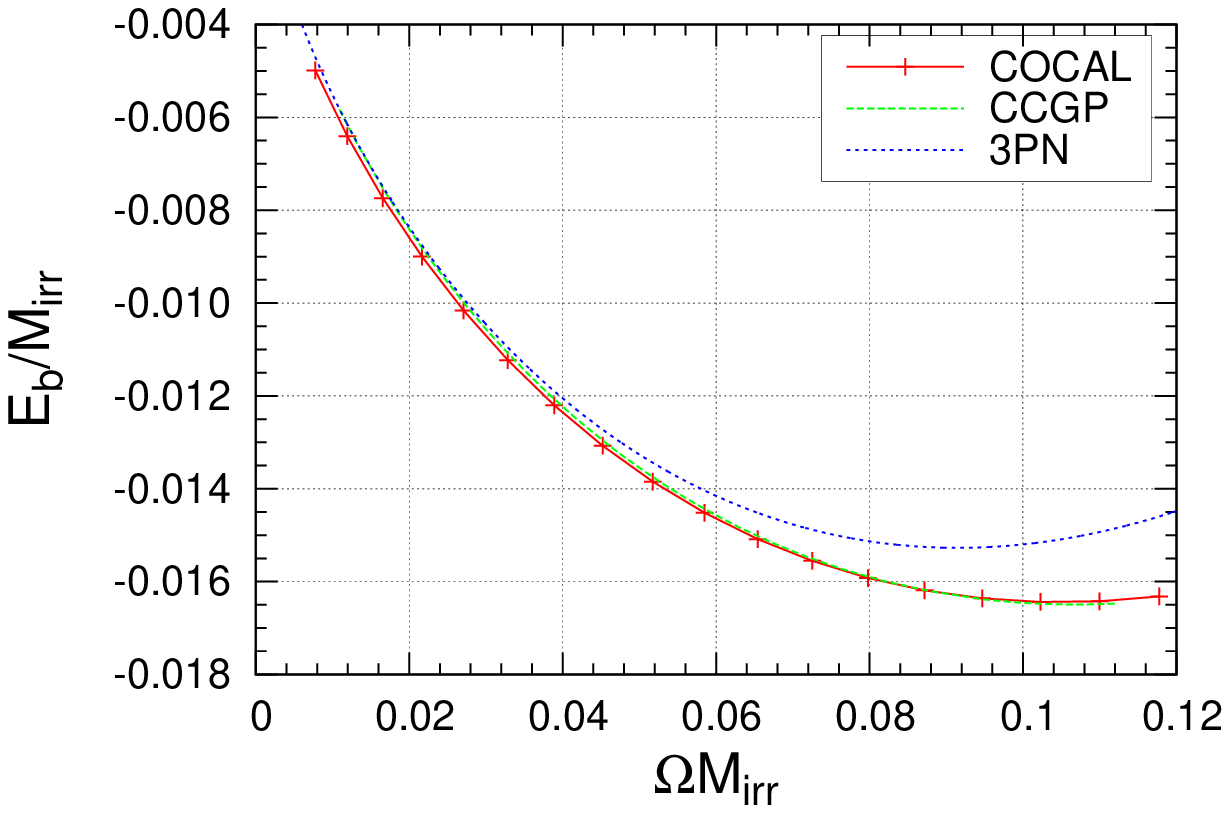}
\includegraphics[height=60mm]{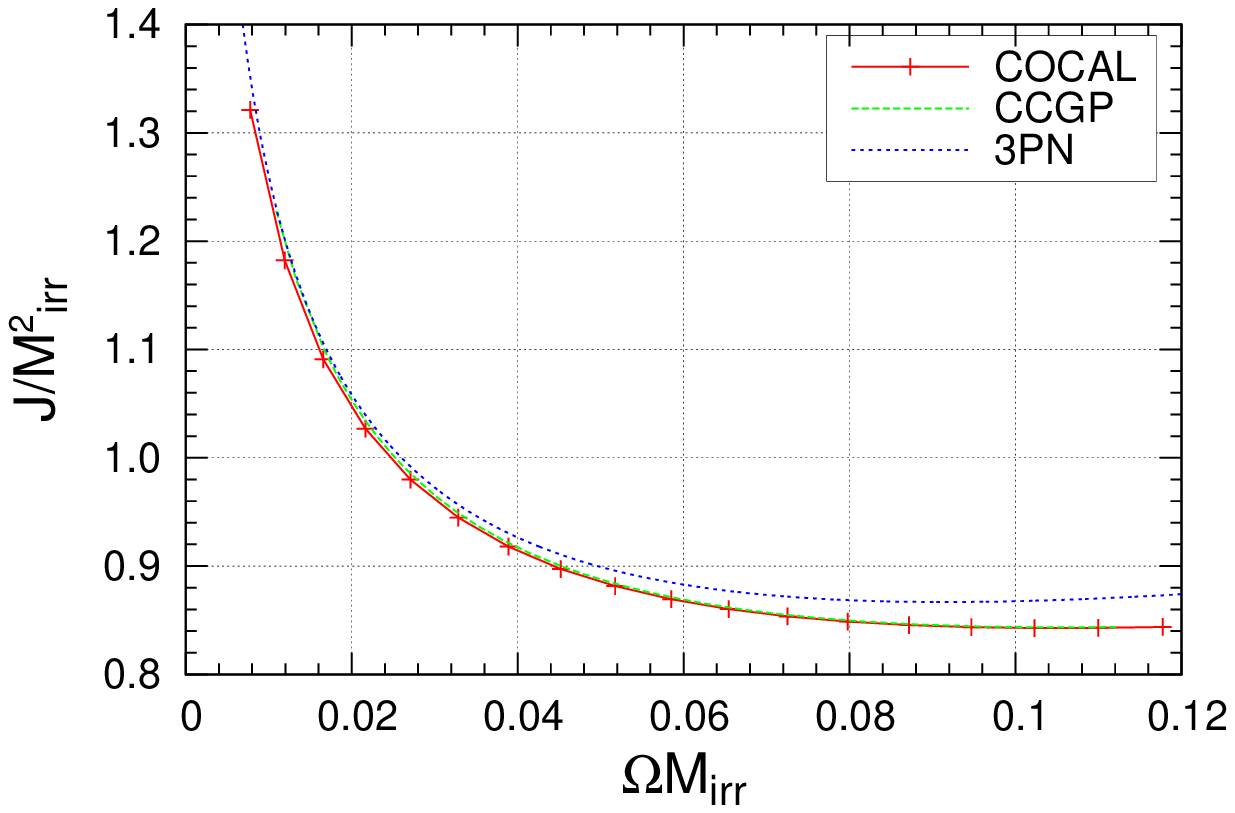}
\caption{Sequence of solutions for conformally flat BBH initial data 
with corotating spins are plotted.  The sequence computed from 
\cocal code is compared with those presented in Caudill et.al. 
\cite{Caudill:2006hw} (CCGP) and with the third post-Newtonian (3PN) 
results \cite{PN}. 
}
\label{fig:BBH_sequence_corot}
\end{center}
\end{figure}

\subsection{Comparison of the circular solutions of BBH initial data for 
\kadath and \cocal} \label{sec:comparison}

Following \cite{GGB}, we obtain the angular velocity $\Omega$ of 
a circular orbit of BBH initial data from an assumption that 
an equality of ADM mass and Komar mass, $\Madm = \MK$, 
is satisfied for the circular orbit, where $\Madm$ and $\MK$ are 
defined by 
\beqn
\Madm &=& \frac{1}{16\pi}\int_{S_\infty} (f^{ac}f^{bd}-f^{ab}f^{cd})\pa_b \gamma_{cd}dS_a
\nonumber\\
&=& -\frac{1}{2\pi}\int_{S_\infty} D^a \psi dS_a, 
\\
\MK &=& -\frac{1}{4\pi}\int_{S_\infty} \na^\alpha t^\beta dS_{\albe}
\nonumber\\
&=& \frac{1}{4\pi}\int_{S_\infty} D^a \alpha dS_a. 
\eeqn
In \cocal, the surface integrals 
are calculated at a certain finite radius, typically around $r\sim(10^4M)$. 
 
In \kadath, these integrals are usually evaluated at spatial infinity 
as its definition, 
$\displaystyle 
\int_{S_\infty} := \lim_{r\rightarrow \infty} \int_{S_r} 
$ 
with $S_r$ the sphere of radius $r$.  
However, in the case of a large separation,
it turned out that this was not giving good results. 
The precision of the results, measured by convergence of the value 
of $\Omega$, is much better when one demands that, at spatial infinity,  
\beq
1 - \alpha\Psi^2 \rightarrow  \mathcal{O}\left(r^{-2}\right).
\eeq
The difference between the Komar and ADM mass is then of the order of $10^{-4}$ for 
the higher resolution, thus giving a measure of the overall error of the code.

For a converged circular solution, we also calculate 
ADM angular momentum: 
\beq
J = -\frac{1}{8\pi}\int_{S_\infty} \Kab \phi^b dS_a.  
\eeq
The above quantities are normalized by the irreducible mass 
of AH, $\Mirr$, which is defined from the surface area of AH, 
namely the area integral over the BH excision surface $S_a$, 
\beq
\AHarea = \int_{S_a} \psi^4 r_a^2 d\Omega.  
\eeq
We write ${\Mirr}_{1} :=\sqrt{{\AHarea}_{1}/16\pi}$ 
and ${\Mirr}_{2} :=\sqrt{{\AHarea}_{2}/16\pi}$ for each BH, 
and write $\Mirr = {\Mirr}_{1} + {\Mirr}_{2}$ for a total mass.  

In Table \ref{tab:BBH_comparison_D12D30}, 
global quantities normalized by $\Mirr$ are presented for  
the irrotational BBH data computed from \kadath and \cocal 
at the separations $d_s/r_a = 12$ and  $d_s/r_a = 30$. 
Three different resolutions for \kadath are given, mainly $11$, $13$ and 
$15$ points in each dimension, and two (lower and higher) resolutions, 
H2 and H3, are used for the computations of \cocal.  The relative
differences between the highest resolution results from both codes are 
also indicated.  The convergence of these quantities is plotted in 
Fig.~\ref{fig:conv} for both separations, $d_s/r_a = 12$ (left panels) 
and  $d_s/r_a = 30$ (right panels).  
All plots indicate a nice convergence of the global quantities, 
when the resolutions are increased.



In Fig.\ref{fig:plot_errors}, plotted are fractional errors of 
metric potentials of the same BBH data between those calculated 
from \kadath and \cocal.  
The errors are defined by 
$100\times|q(\mbox{\sc cocal}) - q(\kadath)|/|q(\kadath )|\,[\%]$, 
and those of the conformal factor $\psi$, lapse $\alpha$, 
and $y$-component of the shift $\beta_y$ are plotted 
along the $x$-axis which intersects with the centers of two BHs
for the case with the separation $d_s/r_a=12$.  Resolutions 
are H3 for \cocal and 15 points for \kadath.  
As seen from the figure, the metric potentials from the two 
codes agrees well.  Note that the relatively large error 
in $\beta_y$ near $x/r_a \approx 6$ is due to $\beta_y$ 
crossing zero, and hence its fractional error diverges there.

\subsection{Solution sequence for corotating BBH data}
\label{sec:BBHcorot}

Finally, we present a sequence of solutions for the conformally flat 
BBH initial data computed from \cocal.  In this computation, the boundary 
condition of shift $\beta^i$ (\ref{eq:AHbetaBC})
is replaced by that for the BBH with corotating spins, 
\beq
\left.\GB^i\right|_{r=r_a} \,=\, \frac{n_0}{\GC^2}s^i 
- \Omega\,\phi^i_{\rm cm},
\label{eq:AHbetaBC_corot} 
\eeq
where the vector $\phi^i_{\rm cm}$ is a generator of rotation 
around the center of mass whose components in Cartesian coordinates 
is written $\phi^i_{\rm cm}=(-y_{\rm cm},x_{\rm cm},0)$.  
The orbital angular velocity parameter $\Omega$ is evaluated 
with the same method as mentioned in the previous section.  

In Fig.~\ref{fig:BBH_sequence_corot}, a sequence of the corotating 
BBH data computed from \cocal with resolution H3 is compared with 
those of the paper by Caudill et.al.~\cite{Caudill:2006hw}, and of the 
third post-Newtonian (3PN) circular orbits \cite{PN}.  
In the top panel, the binding energy normalized by the irreducible 
mass $E_b/M_{\rm irr}:=(\Madm-M_{\rm irr})/M_{\rm irr}$ is plotted 
against the normalized angular velocity $\Omega M_{\rm irr}$, 
and similarly in the bottom panel, the angular momentum $J/M^2_{\rm irr}$.  
The curves calculated from \cocal agree well with the other two curves.  
In the smaller $\Omega M_{\rm irr}$ (large separation), the curves 
from the \cocal are slightly smaller than those of the other works.  
The size of this error is comparable to that listed in Table 
\ref{tab:BBH_comparison_D12D30}.  The error in $E_b/M_{\rm irr}$ 
for the model with $d_s/r_a=30$ is around $-1.2\%$ whose separation 
corresponds to the one in Fig.\ref{fig:BBH_sequence_corot}
with $\Omega M_{\rm irr} \sim 0.026$ ($d_s/r_a =31.25$).  
The data used in Fig.~\ref{fig:BBH_sequence_corot} are 
tabulated in Table.~\ref{tab:corot_BBH_sequence}.

For computing a solution with the \cocal code, each iteration 
takes about 3 minutes for H3 resolution with a single CPU (1 core) 
of Xeon X5690 3.46GHz, and each run uses about 6GB of RAM.  
A convergence to a circular solution is achieved after 500-700 
iterations, during which an iterative search for the circular 
$\Omega$ is made 5-7 times.  

%
\begin{table*}
\begin{tabular}{cccccccc}
\hline
$r_a$ &  $d_s/r_a$ &  $d_s/\Mirr$ & $\Omega \Mirr$ & 
$\Madm/\Mirr$ & $E_b/\Mirr$ & $J/\Mirr^2$ & $\Mirr$ \\
0.03 &  83.333 & 24.327 & 0.0077690 & 0.99501 & -0.0049873 & 1.3211 & 0.10277 \\
0.04 &  62.500 & 18.126 & 0.011925  & 0.99360 & -0.0064032 & 1.1824 & 0.13793 \\
0.05 &  50.000 & 14.403 & 0.016575  & 0.99226 & -0.0077416 & 1.0910 & 0.17357 \\
0.06 &  41.667 & 11.921 & 0.021647  & 0.99100 & -0.0089952 & 1.0268 & 0.20972 \\
0.07 &  35.714 & 10.147 & 0.027086  & 0.98984 & -0.010159  & 0.97988 & 0.24638 \\
0.08 &  31.250 & 8.8151 & 0.032848  & 0.98877 & -0.011228  & 0.94470 & 0.28360 \\
0.09 &  27.778 & 7.7785 & 0.038898  & 0.98780 & -0.012201  & 0.91792 & 0.32140 \\
0.10 &  25.000 & 6.9483 & 0.045207  & 0.98693 & -0.013074  & 0.89736 & 0.35980 \\
0.11 &  22.727 & 6.2682 & 0.051746  & 0.98615 & -0.013847  & 0.88156 & 0.39884 \\
0.12 &  20.833 & 5.7007 & 0.058495  & 0.98548 & -0.014518  & 0.86946 & 0.43854 \\
0.13 &  19.231 & 5.2197 & 0.065430  & 0.98491 & -0.015087  & 0.86032 & 0.47896 \\
0.14 &  17.857 & 4.8067 & 0.072533  & 0.98445 & -0.015555  & 0.85355 & 0.52011 \\
0.15 &  16.667 & 4.4481 & 0.079787  & 0.98408 & -0.015922  & 0.84873 & 0.56204 \\
0.16 &  15.625 & 4.1337 & 0.087174  & 0.98381 & -0.016191  & 0.84550 & 0.60479 \\
0.17 &  14.706 & 3.8557 & 0.094679  & 0.98364 & -0.016362  & 0.84360 & 0.64839 \\
0.18 &  13.889 & 3.6081 & 0.10229   & 0.98356 & -0.016439  & 0.84279 & 0.69289 \\
0.19 &  13.158 & 3.3860 & 0.10998   & 0.98358 & -0.016424  & 0.84290 & 0.73834 \\
0.20 &  12.500 & 3.1856 & 0.11775   & 0.98368 & -0.016320  & 0.84377 & 0.78477 \\
\hline
\end{tabular}
\caption{
Solution sequence of equal mass BBH with corotating spins.  
In the computation, parameters H3 listed in Table 
\ref{tab:BBHtest_grids} are used with 
a fixed separation $d_s=2.5$, and 
with varying the excision radius $r_a$ of BH.  
}
\label{tab:corot_BBH_sequence}
\end{table*}

\section{Discussion}

We have presented additional convergence tests of \cocal code 
focusing on the BBH data.  Especially the conformally flat 
initial data of BBH in circular orbit calculated from \cocal code 
are compared with those from \kadath code.
We demonstrated that the results from both codes converge toward each other
for large and small separations (see also \cite{TU2012}).  

As fully discussed in \cite{Caudill:2006hw}, 
a corotating sequence presented in Sec.\ref{sec:BBHcorot} is 
not considered as a model for inspiraling BBH, because BBH tides 
do not effectively work to spin up the BH to synchronize the BH spin 
with orbital motion within the time of inspirals.  In \cite{Caudill:2006hw}, 
the authors describe a more realistic sequence of BBH inspiral where the 
spin angular momentum of the AH is conserved.  
We will present elsewhere the performance of the \cocal code 
for computing those sequences to model BBH inspirals, which 
is a necessary step to compute more complex binary systems 
including black hole-neutron star binaries.  
\\
\acknowledgments
This work was supported by 
JSPS Grant-in-Aid for Scientific Research(C) 23540314 
and 22540287, and MEXT Grant-in-Aid for Scientific Research
on Innovative Area 20105004.  
We thank Eric Gourgoulhon for discussion.  
KU thanks Charalampos Markakis, and Noriyuki Sugiyama 
for discussion.  

\end{document}